\documentstyle[twoside,fleqn,espcrc2]{article}

\newcommand{\AmS}{{\protect\the\textfont2
  A\kern-.1667em\lower.5ex\hbox{M}\kern-.125emS}}
\def\bbbr{{\rm I\!R}}

\title{Universal amplitude ratios in finite-size scaling: three-dimensional
  Ising model}

\author{Martin Weigel 
        and Wolfhard Janke\address{Institut f\"ur Theoretische Physik,
        Universit\"at Leipzig,
        Augustusplatz 10/11, 04109 Leipzig, Germany}}
       
\begin{document}

\begin{abstract}
  Motivated by the results of two-dimensional conformal field theory (CFT) we
  investigate the finite-size scaling of the mass spectrum of an Ising model on
  three-dimensional lattices with a spherical cross section. Using a
  cluster-update Monte Carlo technique we find a linear relation between the
  masses and the corresponding scaling dimensions, in complete analogy to the
  situation in two dimensions. Amplitude ratios as well as the amplitudes
  themselves appear to be universal in this case.
\end{abstract}

\maketitle

\section{INTRODUCTION}

Divergent correlation lengths of a statistical system approaching a critical
point establish the symmetry of scale invariance on which finite-size scaling
theory is based. In a continuum critical theory, however, there are additional
symmetries present, namely translational, rotational and inversion invariance.
The implications of the group of these symmetries, as explored in conformal
field theory \cite{cardy:domb,henkel:book}, go far beyond simple scaling theory,
especially for two-dimensional systems, where the whole content of scaling
operators as well as for example higher correlation functions can be generically
extracted from this symmetry.

As for a prominent result, conformal invariance suffices to determine the
critical two-point function in the plane. Using a conformal (logarithmic)
transformation that wraps the plane around an infinite-length cylinder one thus
arrives at an expression for the correlation function of a primary (conformally
covariant) operator on the cylinder \cite{cardy:84a}, a geometry one generically
considers in transfer matrix calculations. The fact that this transformation
does only affect the radial but not the angular part of the coordinates led
Cardy \cite{cardy:85a} to a generalization for the higher-dimensional geometries
$S^{d-1}\times\bbbr$. Thus, exactly as in two dimensions, the large distance
exponential decay of correlations is determined by a correlation length that is
given by the inverse of the scaling dimension $x$ of the operator under
consideration,
\begin{equation}
  \xi=\frac{R}{x},
  \label{conj}
\end{equation}
where $R$ is the radius of $S^{d-1}$. It has to be noted, however, that the
meaning of a {\em primary} operator is a priori not well defined in three
dimensions\footnote{For an ansatz for its definition see \cite{lang:93a} and
  references therein.} so that this result has to be considered a conjecture and
a numerical analysis is more than an exercise in this case.

A first attempt to establish such evidence using the Hamiltonian formulation of
the Ising model and Platonic solids as discretization of the sphere $S^2$ was
inconclusive due to the restricted number and size of regular polyhedra
\cite{alcaraz:87a}. Using {\em approximate} quadrangulations we are able to
check Cardy's conjecture for systems sufficiently large to carry out a proper
finite-size scaling (FSS) analysis.

\section{LATTICE DISCRETIZATION}

There are several straightforward choices for model lattices with spherical topology
\cite{lang:96a}; the more natural are variants of a rectangular mesh on a cube.
Different refinements can be applied to reduce the concentration of the
curvature around the corners of the cube such as filling in triangles instead of
the cube corners or projecting the cube on the sphere by varying the link
lengths, i.e., applying an appropriate site-dependent weight function to the
action.  For those cube-like lattices, however, differences in the scaling
behavior of bulk quantities are found to be quite small \cite{lang:96a}.  As for
our observables, there is some evidence to believe that {\em ratios} of
correlation lengths of primary operators are universal
\cite{henkel:87a,yuri:97a,prl:99a}, so that we can expect good agreement
regardless of the lattice used if Cardy's conjecture holds. For the amplitudes
themselves one might find effects of the non-uniformity of the lattice, but would
not expect them to become visible until a very high level of precision is
reached. We thus here use the simple cubic approximation to the sphere by six
$L\times L$ square lattices with suitably chosen boundary conditions; effects of
the discretization will be considered elsewhere \cite{prep}.

Having chosen a lattice approximation to the sphere one has to fix a definition
of the radius $R$ of the sphere the cube with edge length $L$ should correspond
to.  As is easily checked the three possibilities of assigning a unit area to
each lattice site, each pair of bonds, or each lattice square, respectively,
lead to total lattice areas of
\begin{equation}
A = \left\{
\begin{array}{c@{}ll}
6L & (L-2)+8 & {\rm ``sites"}, \\
6L & (L-2)+6 & {\rm ``bonds", ``squares"},
\label{radii}
\end{array}
\right.
\end{equation} 
and generate via the relation $R=\sqrt{A/{4\pi}}$ two different sorts of pseudo
radii, which only differ by a constant shift, thus leading to slightly different
approaches to the leading FSS amplitude.

\section{SIMULATIONAL DETAILS}

We consider a classical, ferromagnetic, nearest-neighbor Ising model with
Hamiltonian
\begin{equation}
  {\cal H}=-J\sum_{\langle i,j\rangle}s_i s_j,\;\;\; s_i=\pm 1
\end{equation}
on lattices compound of the above described cubic discretizations of the sphere
times a linear direction of length $L_z$, modelling the $\bbbr$-direction. In
order to minimize the effect of a finite $L_z$ it was (self-consistently) chosen
so that $L_z/\xi\approx 15$ and periodic boundary conditions in $z$-direction
were applied. Simulations were done using the Wolff single-cluster update algorithm
at an inverse temperature $\beta=0.221\,654\,4(3)$ \cite{talapov:96}, checking
for the influence of a temperature shift off the critical point by reweighting.
To enable a proper FSS analysis system sizes ranging from $L=4$ to $L=12$ were
used, corresponding to radii from about $2$ to $8$ and overall lattice volumes
of $6\,000$ to $300\,000$ spins.

Since the densities of energy and magnetization are (the only) primary operators
in the {\em two-dimensional} Ising model and the lowest lying states of the two
sectors, we here consider the amplitudes of the corresponding (exponential)
correlation lengths $A_{\sigma/\epsilon}=\xi_{\sigma/\epsilon}/R$ and their
ratio. Simulational measurements were done for the (connected) correlation
functions $G^c(z)$ of these operators using a zero mode projection. Estimators
for the correlation lengths that eliminate additive and multiplicative constants
in the exponential decay are then given by
\begin{equation}
  \xi(z)=\Delta{\left[\ln\frac{G^c(z)-G^c(z-\Delta)}{G^c(z+\Delta)-G^c(z)}\right]}^{-1},
\label{diffmethoddelta}
\end{equation}
with $\Delta=1,2,3,\ldots$,
from which overall estimates $\xi(R)$ are formed by averaging. A jackknife
technique was used to reduce estimator bias and for the error analysis; for
details see \cite{prl:99a,pils:98}.

\section{NUMERICAL RESULTS}

\begin{figure}[t]
\begin{picture}(150,145)
    \put(0, 0){\includegraphics{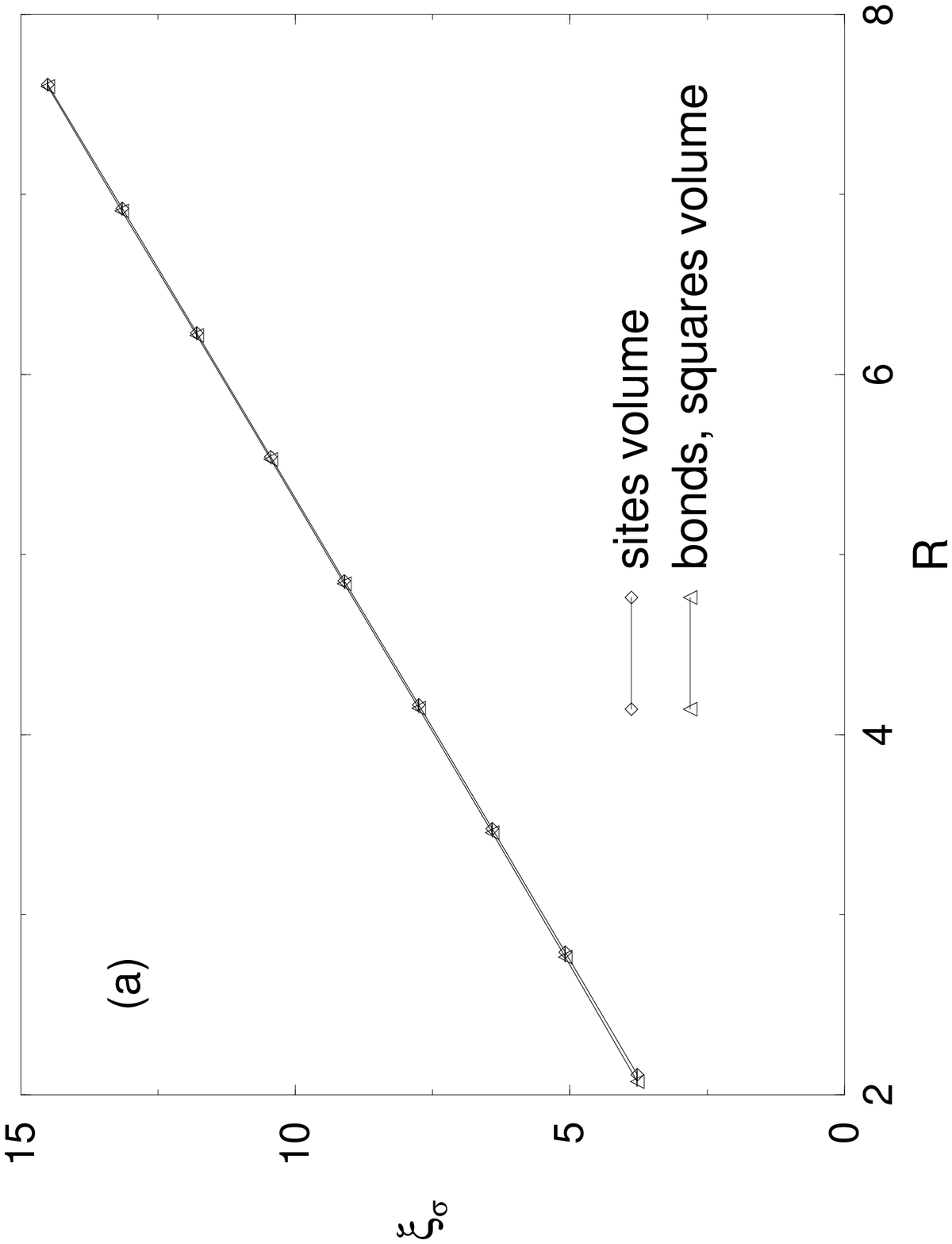}}
  \end{picture}
\begin{picture}(150,145)
    \put(0, 0){\includegraphics{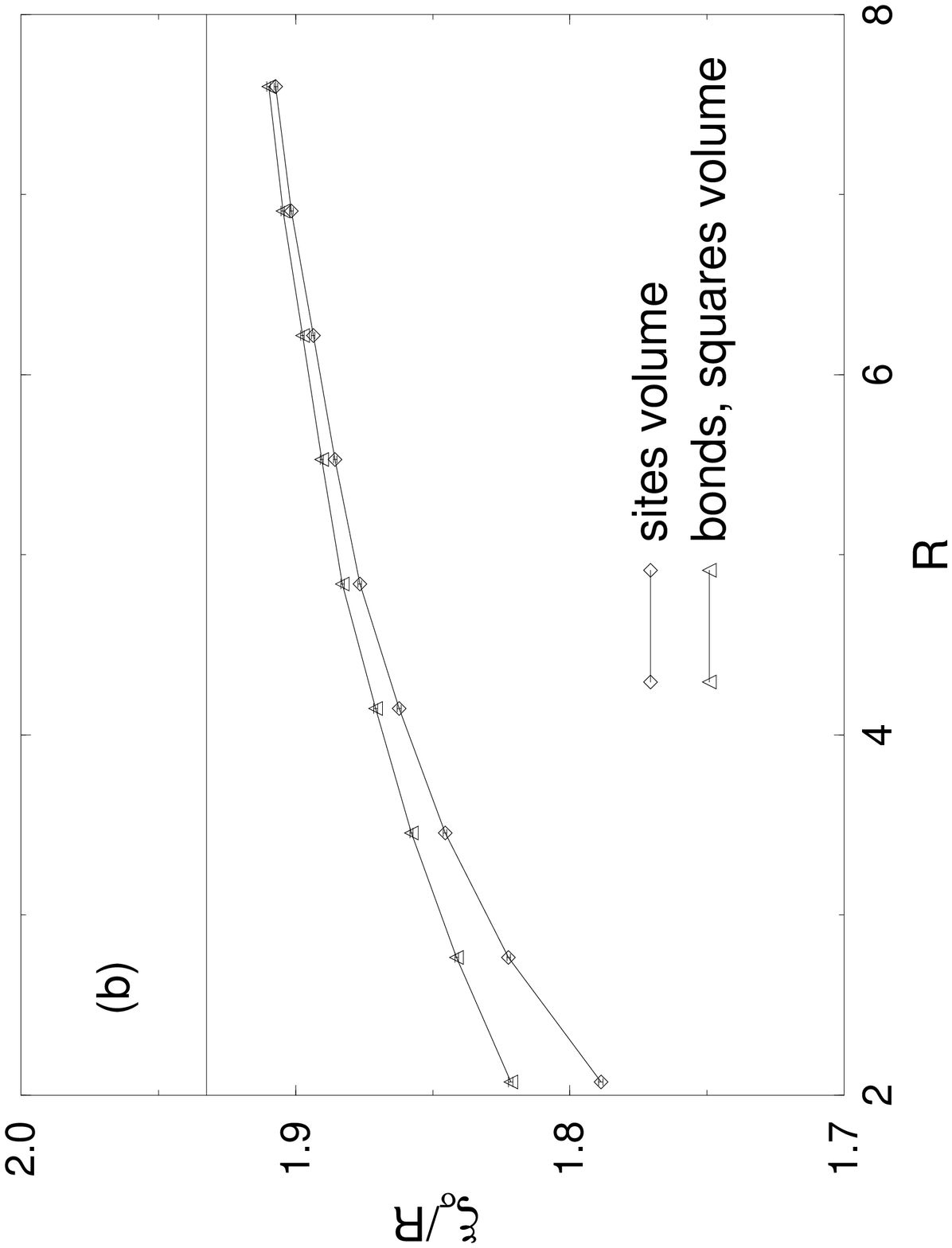}}
  \end{picture}
\caption{(a) FSS plot for the spin correlation length $\xi_\sigma(R)$.
(b) Scaling of the amplitudes $\xi_\sigma/R$. The horizontal line indicates the
conjectured amplitude according to Eq.\,(\ref{conjamp}).}
\label{fig}
\end{figure}

Using the above described procedures one ends up with scaling plots for the
correlation lengths as shown in Fig.\,\ref{fig}(a) for the magnetization density
that show an almost perfect linear behavior and invisible differences between
the different radii definitions according to Eq.\,(\ref{radii}). Plotting the
amplitudes $\xi/R$ (Fig.\,\ref{fig}(b)), however, reveals the presence of
corrections to scaling that enforce the use of nonlinear fits of the form
\begin{equation}
  \xi(R)=AR+BR^{\alpha},
\end{equation}
with a free parameter $\alpha$ that introduces a systematic error to the final
amplitudes $A_{\sigma/\epsilon}$ due to higher order corrections. A combination
of the fits for the two sorts of radii gives the amplitudes
\begin{equation}
  \begin{array}{rcl}
    A_\sigma & = & 1.996(20), \\
    A_\epsilon & = & 0.710(38),
  \end{array}
\end{equation}
which agree well with the conjectured amplitudes
\begin{equation}
  \begin{array}{rcccl}
    A_\sigma^{\rm conj} & = & 1/x_\sigma& = & 1.9324(19), \\
    A_\epsilon^{\rm conj} & = & 1/x_\epsilon & = & 0.70711(35),
  \end{array}
\label{conjamp}
\end{equation}
taking into account the systematic error due to the different radii
definitions that is of the same order of magnitude as the statistical error. The
amplitude ratios compare as:
\begin{equation}
  \begin{array}{ccccl}
  A_\sigma/A_\epsilon & = & & & 2.81(15), \\
  A_\sigma^{\rm conj}/A_\epsilon^{\rm conj} & = & x_\epsilon/x_\sigma & = & 2.7326(16). 
  \end{array}
\end{equation}

\section{CONCLUSIONS}

The presumably universal ratio of the amplitudes of the spin and energy
correlation lengths of the Ising model on a cubic model of $S^2\times\bbbr$
agrees with the inverse ratio of the scaling dimensions as conjectured by Cardy
for the continuum case. Moreover, corrections to scaling due to the non-uniform
distribution of curvature over the lattice that might influence the amplitudes
themselves seem not to be very important at the given level of accuracy; only
the slightly shifted result for the spin amplitude $A_\sigma$ might indicate the
onset of such effects. In connection with the results for systems with toroidal
cross section \cite{prl:99a,pils:98} this seems to indicate a deeper analogy
between the 2D and 3D situations. In this latter case, however, the system is
not conformally related to a flat space which renders impossible an analytical
treatment of Cardy's kind. In the spherical case, on the other hand, conformal
flatness is fulfilled, but the algebraic meaning of the primarity of an operator
remains unclear. Thus our result shows that spin and energy densities of the 3D
Ising model are primary operators, taking Eq.\,(\ref{conj}) as a definition of
primarity in three dimensions.

MW thanks the Deutsche Forschungsgemeinschaft for support through the
Graduiertenkolleg ``Quantum Field Theory".

\end{document}